\documentclass[aps,prresearch,twocolumn,amsmath,amssymb,longbibliography,superscriptaddress]{revtex4-2}% APS journal style
\usepackage{amsmath,amssymb,amsfonts,bm}
\usepackage{xcolor}
\usepackage[most]{tcolorbox}
\usepackage{bbold}
\usepackage{graphicx}% Include figure files
\usepackage{dcolumn}% Align table columns on decimal point
\usepackage{epstopdf}
\usepackage{epsfig}
\usepackage[caption=false]{subfig}
\usepackage[colorlinks=true, breaklinks=true]{hyperref}

\usepackage{braket}
\usepackage{overpic}

\usepackage{enumitem}

\usepackage{multirow}
\usepackage{booktabs}
\usepackage{pifont}

\usepackage{colortbl}

\usepackage{comment}

\def\Irr{\operatorname{{Irr}}}
\newcommand{\rrangle}{\rangle\hspace{-0.8mm}\rangle}
\newcommand{\llangle}{\langle\hspace{-0.8mm}\langle}

\newcommand{\kket}[1]
{|{#1}\rrangle }
\newcommand{\bbra}[1]
{\llangle {#1}| }

\newcommand{\bbraket}[1]{\llangle {#1} \rrangle}

\newcommand{\bc}{\mathrm{bc}}

\newcommand{\GinVar}{\nu}
\newcommand{\permstate}{\sigma}

\usepackage{float}
\newcommand{\Tr}{\operatorname{Tr}}

\newcommand{\NN}{N_{\mathrm{Th}}}

\newcommand{\wg}{\mathrm{Wg}}

\newcommand{\tr}[1]{\mathrm{tr}[#1]}

\newcommand{\be}{\begin{equation}}
\newcommand{\ee}{\end{equation}}
\newcommand{\ba}{\begin{aligned}}
\newcommand{\ea}{\end{aligned}}

\newcommand{\bmult}{\begin{multline}}
\newcommand{\emult}{\end{multline}}

\newcommand{\red}[1]{{\color{red} #1}}

\usepackage{ifthen}

\newcommand{\Andrea}[2][]{%
  \textcolor{orange}{#2}%
  \ifthenelse{\equal{#1}{}}{}{%
    \textcolor{orange}{\textbf{[Andrea: #1]}}%
  }%
}

\newcommand{\nf}{n_{\mathrm{F}}}

\newcommand{\HH}{\mathcal{N}}

\def\Lth{L_{\rm Th}}

\graphicspath{{Figures/}}

\begin{document}
\def\titleinfo{Universal distributions of overlaps from   generic dynamics  \\ in quantum many-body systems}

\title{\titleinfo} 

\author{Alexios Christopoulos}
\email{alexios.christopoulos@cyu.fr}
\affiliation{Laboratoire de Physique Th\'eorique et Mod\'elisation, 
~CNRS UMR 8089, CY Cergy Paris Universit\'e, \\ F-95302 Cergy-Pontoise, France}

\author{Amos Chan \hphantom{$^\dag$}}
\email{amos.chan@warwick.ac.uk}
\affiliation{Department of Physics, University of Warwick, Coventry, CV4 7AL, United Kingdom}
\address{Department of Physics, Lancaster University, Lancaster LA1 4YB, United Kingdom}

\author{Andrea De Luca \hphantom{$^{\dag \dag}$}}
\email{ andrea.de-luca@cyu.fr}
\affiliation{Laboratoire de Physique Th\'eorique et Mod\'elisation, 
~CNRS UMR 8089, CY Cergy Paris Universit\'e, \\ F-95302 Cergy-Pontoise, France}

\date{\today}

\begin{abstract}
We study the distribution of overlaps with the computational basis of a quantum state generated under generic quantum many-body chaotic dynamics, without conserved quantities, for a finite time $t$.
We argue that, scaling time logarithmically with the system size $t \propto \log L$, the overlap distribution converges to a universal form in the thermodynamic limit, forming a one-parameter family that generalizes the celebrated Porter-Thomas distribution. The form of the overlap distribution only depends on the spatial dimensionality and, remarkably, on the boundary conditions. This picture is justified in general by a mapping to Ginibre ensemble of random matrices and corroborated by the exact solution of a random quantum circuit. 
Our results derive from an analysis of arbitrary overlap moments, enabling the reconstruction of the distribution.
Our predictions also apply to Floquet circuits, i.e., in the presence of mild quenched disorder.
Finally, numerical simulations of two distinct random circuits show excellent agreement, thereby demonstrating universality.
\end{abstract}

\maketitle

\section{Introduction} 
Quantum many-body system dynamics, particularly their scrambling capabilities and implications for quantum chaos and holography, have been intensively studied~\cite{Calabrese_2016,PhysRevX.7.031016,vonKeyserlingk2017a, cdc1, Cotler_2017, Hayden_2007}.
The concept of quantum $k$-state designs, which mimics the Haar distribution's uniformity in a Hilbert space, has emerged as pivotal~\cite{DiVincenzo_2002, ambainis2007quantum, Gross_2007, Roberts_2017}. These designs are explored through unitary and non-unitary processes~\cite{harrow2009random, Brand_o_2016, chan2024projected, ho2022exact, Ippoliti:2022bsj, PhysRevA.107.032215} and have broad implications from quantum computing to theories of black holes~\cite{Boixo_2018, arute2019google, susskind2014computational, Brown_2018, haferkamp2022linear, brandao2021complexity, Hayden_2007, Sekino_2008, Shenker_2014, Maldacena_2016}.
 Understanding how many operations are needed to achieve a good quantum state design is an open question in many cases, with important applications from quantum computation, particularly benchmarking~\cite{Boixo_2018, arute2019google}, to theories of black holes~\cite{susskind2014computational, Brown_2018, haferkamp2022linear, brandao2021complexity, Hayden_2007, Sekino_2008, Shenker_2014, Maldacena_2016}.
 In this regard, Random Unitary Circuits (RUCs) have served as vital toy models in quantum information and many-body physics, capturing numerous universal features for strongly coupled quantum dynamics~\cite{doi:10.1146/annurev-conmatphys-031720-030658}. Recent explorations involve the use of RUCs to examine operator growth \cite{PhysRevX.8.021014,vonKeyserlingk2017a, PhysRevX.8.021013, Huse2017} and entanglement \cite{PhysRevX.7.031016,PhysRevB.98.035118,PhysRevB.99.174205} amid chaotic evolution, and spectral statistics~\cite{cdc1, cdc2, friedman2019, Kos_2018, bertini_exact_2018}. 

\begin{figure}[t]
  \begin{overpic}[width=0.48\textwidth, trim={0 1.68cm 0 0},clip]{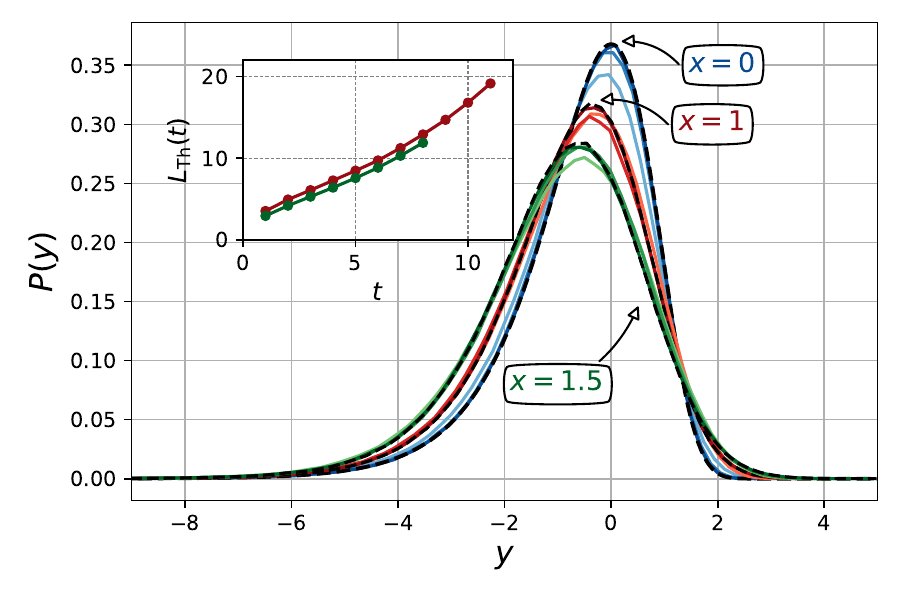}
    \put(4,52){(a)}\put(17,9){\begin{tcolorbox}[
    colframe=black,          % Frame color (black)
    colback=gray!10,         % Background color (10% grey)
    rounded corners,         % Rounded corners for the box
    boxrule=0.5mm,           % Frame thickness
    width=0.22\linewidth,    % Width of the box
    enhanced,                % Enable advanced features
    opacityback=1,           % Full background opacity
    halign=center,           % Ensures text is horizontally centered
    valign=center,            % Ensures text is vertically centered
    overlay
]
RPM pbc
\end{tcolorbox}}
  \end{overpic}
  \begin{overpic}[width=0.48\textwidth, trim={0 0.3cm 0 0.4cm},clip]{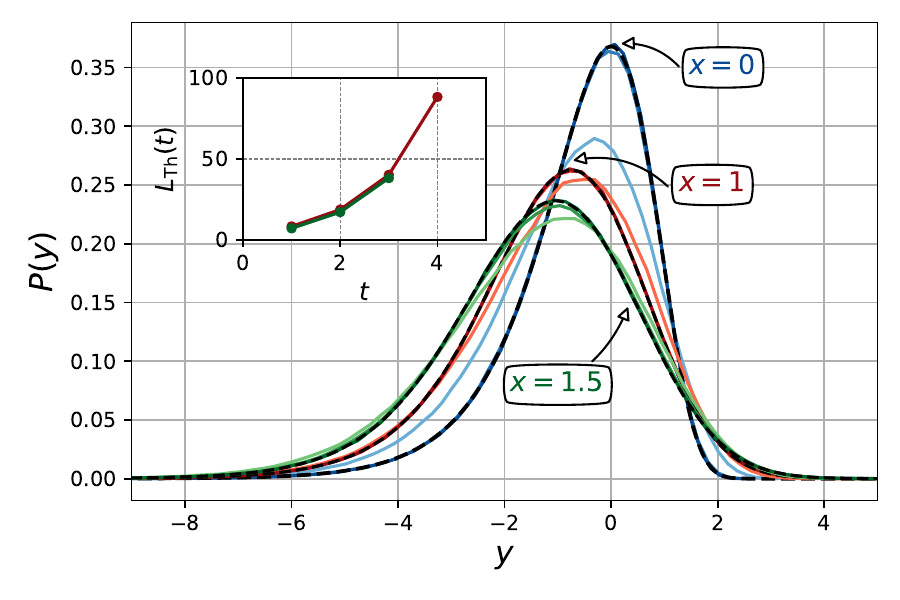}
    \put(4,60){(b)}\put(17,18){\begin{tcolorbox}[
    colframe=black,          % Frame color (black)
    colback=gray!10,         % Background color (10% grey)
    rounded corners,         % Rounded corners for the box
    boxrule=0.5mm,           % Frame thickness
    width=0.22\linewidth,    % Width of the box
    enhanced,                % Enable advanced features
    opacityback=1,           % Full background opacity
    halign=center,           % Ensures text is horizontally centered
    valign=center,            % Ensures text is vertically centered
    overlay
]
BWM obc
\end{tcolorbox}}
  \end{overpic}
\caption{Comparison of the distribution of $y = \log w$ between numerical simulation and the theoretical prediction (black dashed line) for different values of $x$ and increasing value of the time $t$, which is indicated with darker shades of the same colour. For each $t$, the value of $L \sim \Lth(t)$ (shown in the insets) is chosen so that $\mathbb{E}[y]$ matches the theoretical prediction. The sub-figure (a) demonstrates the results from the pbc, numerical simulation of  RPM at $q=2,\epsilon=1$, whereas sub-figure (b); the obc, numerical simulation for a BWM where the local 2-site gate is chosen independently from the Haar distribution for $q=2$. For more details, see Appendixes ~   \ref{app:models},\ref{app:NumericalSim}.}
\label{fig:pdfbc}
\end{figure}

In this paper, we study how quantum states generated by generic quantum circuits expand in a computational basis of reference. For concreteness, consider a system with $N$ quqits and a quantum state $\ket{\Psi} =W(t) \ket{\Psi_0}$ with $W(t)$ a quantum circuit of depth (or time) $t$ drawn from an ensemble of statistically similar circuits without conserved quantities, acting on a factorized reference state $\ket{\Psi_0}$. We denote states in the computational basis as $\ket{\textbf{a} = a_1, \ldots, a_L}$ with $a_i=0,1,\ldots,q-1$ and choose for definiteness $\ket{\Psi_0} = \ket{\textbf{a} = 0,\ldots,0}$.  Quantum dynamics as $t$ increases will produce an increasingly delocalized wave function in Hilbert space.  To quantify this effect, it is natural to analyze the (normalized) overlaps $w_{\textbf{a}} = \mathcal{N} |\braket{\textbf{a}|\Psi}|^2$, where $\mathcal{N} = q^N$ is the Hilbert space dimension. The empirical overlap distribution $\rho_{\ket{\Psi}}(w) = \mathcal{N}^{-1}\sum_{\textbf{a}} \delta(w - w_\textbf{a})$ is the quantity of interest in this work. 
Other commonly used measures of this spreading in Hilbert space include the participation entropy and inverse participation ratio~\cite{mace2019multifrac, B_cker_2019, Turkeshi_2023, thurkeshi2024partentropy,PhysRevLett.134.050405} and can be easily reconstructed from $\rho_{\ket{\Psi}}(w)$.
\begin{figure}[ht]
    \centering    \includegraphics[width=0.9\linewidth]{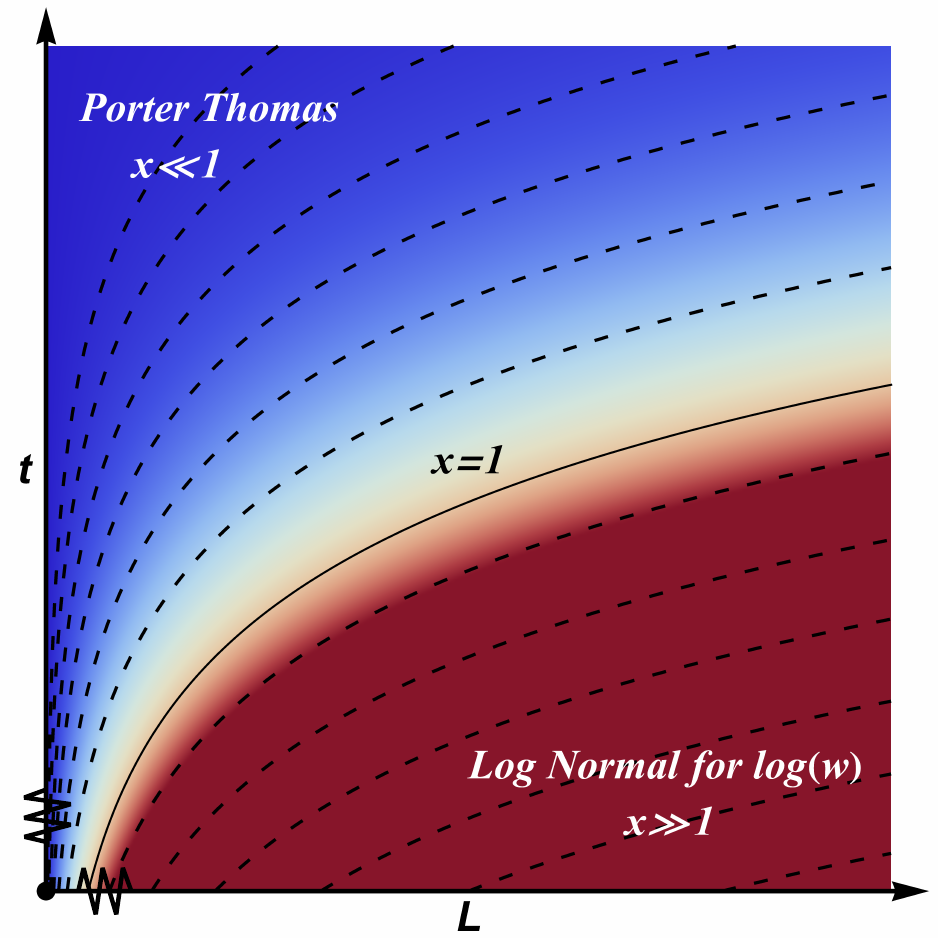}
    \caption{The space-time of our problem parametrized as $(x,t)$ according to $L=x L_{\text{Th}}(t)$ coming from the definition of $x$. The variable $x$ defines a family of coordinate space-time curves, as indicated by the dashed lines. The solid curve represents the curve of $x=1$ and separates the regions of $x<1, x>1$, above and below, respectively. The scaling limit is taken at $t,L \to \infty$, implying that our results for $p(w;x)$ represent its behaviour at the upper right part of the graph. Moreover, $p(w;x)$ has two characteristic limits: in the region of space-time where $x \ll 1$ (blue coloured)  it approaches the  PT distribution, whereas in the region of $x\gg 1$ (red coloured), the distribution of $y=\log w$ approaches a log normal one. }
    \label{fig:spacetime}
\end{figure}
At $t=0$, only one bitstring is populated and $\rho_{\ket{\Psi_0}}(w) \simeq \delta(w)$ at large $N$. Conversely, by increasing $t$, most bitstrings are expected to be populated so that many $\delta$ peaks far from zero appear. In the thermodynamic limit, a continuous distribution can emerge and in the presence of typicality $\rho_{\ket{\Psi}}(w)$ converges weakly to $ \mathbb{E}[\rho_{\ket{\Psi}}(w)] =: \rho(w; t, N)$, where $\mathbb{E}[\ldots]$ denotes the ensemble average at a given depth $t$. For instance, generically $W(t\to\infty)$ provides a sampling of the Haar distribution for which $\rho(w; t, N)$ is known to converge to the Porter-Thomas(PT) distribution, which for $N \to\infty$, takes the form $\rho_{\rm PT}(w) := e^{-w}$~\cite{PhysRev.104.483}.
Studying the crossover from $t=0$ to $t\to\infty$ represents the fundamental question of this paper. An indication is provided by circuit ensembles approximating Haar unitaries for polynomials of degree $k$ (a so-called $k$ design). For a $k$ design, $\mathbb{E}[w_\textbf{a}^{k'}] = (k')!$ for all $k' = 0,\ldots, k$ and already for a 2-design, $w$ cannot be too small, $\mathrm{Prob}(w >\alpha) > (1-\alpha)^2/2$ for all $\alpha \in [0,1]$~\cite{Hangleiter2018}, a phenomenon dubbed \textit{anticoncentration}. The timescales for building  $k$-designs in this context have been studied in~\cite{hunterjones2019}: in the limit of large local dimension, a geometrical interpretation similar to the one proposed for entanglement entropies emerges~\cite{PhysRevB.99.174205, PhysRevX.10.031066}, suggesting that RUCs form approximate unitary $k$-designs at $t \sim O(Nk)$. 
In this work, we provide a complete characterization of the distribution $\rho(w)$ at intermediate $t$ in the thermodynamic limit $N \to \infty$. Formally, we identify a scaling regime governed by a single parameter $x = N/\NN(t)$ (Fig.~\ref{fig:spacetime}), which defines a family of universal distributions $\rho(w; t, N) \to \rho(w; x)$, largely independent of microscopic details, but controlled by spatial dimension ($d=1$ or $d>1$) and boundary conditions (periodic or open). Here  $\NN(t)$ denotes a volume scale within which complete scrambling has occurred.
Its growth is generally exponential with $t$. Indeed, 
in $1$D, $\NN(t) \to \Lth(t)$, a length scale, analogous to the one introduced in~\cite{cdc2, Shivam_2023, chan2021manybody, friedman2019} for the spectral form factor. We derive below that $\Lth(t) \sim e^{S_2(t)}$, with $S_2(t)$ the bipartite second Rényi entropy, known to grow linearly in time from low-entangled initial states~\cite{PhysRevX.7.031016}.  In the scaling limit $L, t \to \infty$ with $x = L/\Lth(t)$ fixed, we obtain exact predictions for the distributions of $w$. In all cases, we express $w$ as the product of two independent random variables $w = w_0 g_x$, where $w_0$ follows PT distribution and $g_x$ encodes the finite-time corrections, with $g_{x \to 0} = 1$ so that PT is recovered for $w$. For open boundary conditions (obc), 
$g_x \to g_x^{\rm obc}$ follows a log-normal distribution and we have the explicit expression
\begin{equation}
\label{eq:pobcw}
p_{\rm obc}(w; x)=  \int_{\mathbb{R}} \frac{du \; e^{-u^2 / 2 + x}}{\sqrt{2\pi}} \exp\left[-w e^{u \sqrt{x}+\frac{3 x}{2}} \right] \,.
\end{equation}
For periodic boundary conditions (pbc), there is no simple formula, but we relate $g_{x}^{\rm pbc}$ to the trace of large random matrix, explicitly
\begin{equation}
\label{eq:pbssample}
    g_{x}^{\rm pbc} \stackrel{\text{\scriptsize{in law}}}{=} \lim_{n \to \infty} \frac{1}{n} \mathrm{tr}\left[ e^{\sqrt{x n} H + x D} \right] \,,
\end{equation}
where $H$ is standard GUE random matrix of size $n$ and $D = \operatorname{diag}(-1/2, -3/2, \ldots, -(2n-1)/2)$. 
We derive these results first by expressing the moments $\mathbb{E}[w^k_{\textbf{a}}]$ diagrammatically (Fig.~\ref{fig:circuit}a) and justifying the mapping of the transfer matrix in the spatial direction to the Ginibre ensemble of random matrices. Through averaging in replica space, moments appear as partition functions of a gas of domain walls, strongly diluted in the scaling limit, justifying universality.
%Then, the emergence of the log-normal component can be traced back to the gaussian fluctuations of the Lyapunov exponent in a product of random matrices. 
Secondly, we perform an exact analysis of the Random Phase Model (RPM) \cite{cdc2} in the limit of large local Hilbert space dimension, also confirming Eqs.~(\ref{eq:pobcw},\ref{eq:pbssample}). Furthermore, we perform numerical simulations in Fig.~\ref{fig:pdfbc} for two different models, substantiating the universality.

The derived theory can be used as a stepping stone to explore different setups as well. In particular, an interesting extension is the Floquet circuits~\cite{cdc1, cdc2}, where the same gates are applied repeatedly over time. In this case, a strong quenched spatial disorder can induce many-body localization~(MBL), suppressing thermalization and scrambling~\cite{MBLreview1, MBLreview2, MBLreview3}, and consequently the generalized PT distribution does not emerge (numerically corroborated in Appendix~ \ref{app:NumericalSim}). At weak disorder, the system can still be in a thermalizing phase but where weak links significantly influence transport~\cite{LuitzReview} and entanglement properties~\cite{PhysRevB.98.035118}. Following the arguments in~\cite{PhysRevB.98.035118}, we expect that weak links will only modify the growth $\log\Lth(t) \propto t^\alpha$, where the exponent $\alpha<1$ changes continuously with the strength of the disorder. That the same distributions also apply to the Floquet case is also confirmed by our simulations away from MBL~\cite{Note1} for the brick-wall model (BWM).

Finally, we mention that projective measurements of all qubits in the computational basis yield a bitstring sampled according to $\operatorname{Prob}(\textbf{a}) = w_\textbf{a}/ \mathcal{N} $. This sampling problem is hard for a classical computer and has therefore been used to exhibit quantum supremacy~\cite{Boixo_2018, arute2019google}. Cross-entropy benchmarking was used to estimate the error, exactly assuming that a perfect implementation of a deep circuit must provide a distribution of probabilities $\operatorname{Prob}(\textbf{a})$ in agreement with Porter-Thomas. Our results provide a benchmark distribution at intermediate depths.

\section{Ginibre ensemble and universality} 
\begin{figure}[h]
    \centering
    \includegraphics[width=0.5\textwidth]{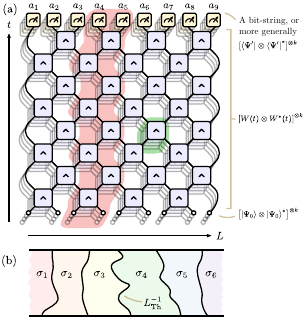}
    \caption{(a) A representation of the powers
    %$w^k = | \braket{\Psi_0 | W'^\dag(t_2=1) W(t_1=3) | \Psi_0}|^{2 k}$,  with $k=2$ for time $t$ and system size $L$.
    $w^k = | \braket{\mathbf{a} = a_1,\dots,a_L | W(t) | \Psi_0}|^{2 k}$,  with $k=2$ for time $t=4$ and system size $L=9$.
    The transfer matrix is highlighted in red. The tensor product,  $u\otimes u^*\otimes \dots \otimes u\otimes u^*$, is highlighted in green, which, upon ensemble-averaging, can be represented as a sum of operators of permutation states. (b) In the Thouless scaling limit, the overlap $\mathbb{E}[w^k]$ can be interpreted as the grand canonical partition function of a dilute gas of domain walls, corresponding to transpositions connecting two permutations and each carrying a fugacity $\Lth^{-1}$. Correspondingly, the size of each domain is $\sim \Lth(t)$. }
    \label{fig:circuit}
\end{figure}
To begin, consider a one-dimensional chain with $N=L$ $q$-dimensional sites. Although the argument applies in a more general form, it is useful to keep in mind a brick-wall RUC $W(t)$ where each gate $u_{i,i+1}(t')$ acting on the sites $i,i+1$ at time $t'$ is chosen independently. Conventionally, a single time step $\Delta t = 1$ contains an even and odd layer. 
We denote with $\mathbb{E}[\ldots]$ the average over the realisation of the circuit. Then, the powers of the overlap $|\braket{\textbf{a}|W(t)|\Psi_0}|^{2k}$ can be represented as in Fig.~\ref{fig:circuit}(a), superimposing $k$ copies of the circuit and its complex conjugate. 
This representation is useful for calculating the moments $\mathbb{E}[w_\mathbf{a}^k]$ that are our goal in deducing the distribution. For each realization of the circuit $W(t)$ and bitstring $\mathbf{a}$, we define a transfer matrix $G_i$ in the spatial direction as the collection of all gates (and initial states) that act in the temporal direction on the $i$-th quqits  (red in Fig.~\ref{fig:circuit}(a)). We denote the product of such transfer matrices by
\begin{equation}
\mathcal{G} = G_1 G_2 \ldots G_L    \;.
\end{equation}
where we omit the dependence on $\mathbf{a}$ to lighten the notation. The overlap $w_{\mathbf{a}}$ can be expressed in terms of the matrix elements of $\mathcal{G}$: for periodic boundary conditions (pbc), one has $w_{\mathbf{a}} = |\Tr[\mathcal{G}]|^2$, while for open boundary conditions (obc), $w_{\mathbf{a}}= |\ell^\dag \mathcal{G} r|^2$ where $\ell, r$ are appropriate boundary vectors whose specific forms are not important. $G_i$ are statistically uncorrelated matrices for different $i$-s, and are of size $M(t) \times M(t)$ with $M(t) = q^{4t -2}$ in the geometry of Fig.~\ref{fig:circuit}(a). The relation between $M(t)$ and $t$ is model-dependent, but the exponential growth is generic. In the following, we omit the time dependence of $M$ unless needed. 
When both $t$ and $L$ are large, we end up with a product of many large matrices, a regime where universality can emerge~\cite{Shivam_2023, deluca2023universality, liu2022lyapunov, PhysRevE.102.052134}. In a coarse-grained picture, we group $\ell$ of these matrices $\tilde{G}_{\mathsf{j}} := G_{\mathsf{j}\ell + 1} G_{\mathsf{j}\ell+1}\ldots G_{(\mathsf{j}+1)\ell}$, with $\mathcal{G} = \prod_{\mathsf{j}=1}^{L/\ell} \tilde{G}_\mathsf{j}$. Under generic many-body chaotic dynamics, the $\tilde{G}_a$'s are non-Hermitian and the natural choice is to assume that, for large enough $\ell$,
ensemble average is equivalent to
sample them from the simplest non-Hermitian random matrices known as the Ginibre ensemble, where all entries of $\tilde{G}_a$ are independently drawn complex Gaussian variables with zero average and variance $\GinVar^2$. In practice, other choices of rotational invariant random matrix ensembles would give the same conclusions in the scaling limit, but the Ginibre ensemble makes the derivation more straightforward. To calculate $\mathbb{E}[w^k_\mathbf{a}]$, we are interested in $k$ copies of $\mathcal{G}$ and $\mathcal{G}^\ast$. From Wick's theorem, we have the identity~\cite{deluca2023universality}
\begin{equation}
\label{eq:Gave}
    \mathbb{E}[\tilde{G}_a \otimes \tilde{G}^\ast_a \otimes \ldots \otimes \tilde{G}_a \otimes \tilde{G}^\ast_a] = \GinVar^{2k} \sum_{\permstate \in S_k} \kket{\permstate} \bbra{\permstate}  \;,
\end{equation}
where, based on the previous assumption, $\mathbb{E}[\ldots]$ now denotes the average over the emergent Ginibre ensemble, and $S_k$ is the symmetric group with $k$ elements. To compactly account for Wick's contractions, we introduce the permutation states $\kket{\permstate} \in \mathbb{C}^{ M^{2k}}$ according to $\llangle \alpha_1,\bar\alpha_1,\ldots, \alpha_k, \bar\alpha_k | \permstate \rrangle = \prod_{q=1}^k \delta_{\alpha_j, \bar\alpha_{\permstate(j)}}$ and $\alpha, \bar\alpha$ are indices for rows/columns of $\tilde G_a$ and $\tilde G^\ast_a$, respectively. Introducing the transfer matrix in the permutation space as $T_{\permstate, \permstate'} = \GinVar^{2k} \bbraket{\permstate | \permstate'}=(\GinVar^{2} M)^{k} M^{- d(\permstate, \permstate')}$ with $d(\permstate, \permstate')$ the transposition distance, we can write $\mathbb{E}[(\mathcal{G}\otimes \mathcal{G}^*)^{\otimes k}] = \sum_{\permstate,\permstate'} [T^{L/\ell - 1}]_{\permstate, \permstate'}\kket{\permstate} \bbra{\permstate'}$ where $\mathcal{G}^*$ is the complex conjugate of $\mathcal{G}$. For large $M \gg 1$, we expand $T(\permstate, \permstate') = (\GinVar^2 M)^k (\delta_{\permstate, \permstate'} + M^{-1} A_{\permstate, \permstate'} + O(M^{-2}))$, where $A_{\permstate, \permstate'}$ is the adjacency matrix of the transposition graph, i.e. it equals one if $\permstate$ and $\permstate'$ differ by one transposition, and vanishes otherwise. 
Introducing the Thouless length as $\Lth(t) \equiv M(t) \ell(t)$,
we define the \textit{Thouless scaling limit} where both $L$ and $t$ are large but $x \equiv L / \Lth(t)$ is kept constant \cite{chan2021manybody}. In this limit, we obtain
\begin{equation}
\label{eq:momentsA}
     \lim_{\substack{t,L \to \infty\\ x= L/\Lth(t) }} 
    \mathbb{E}[w^k_{\mathbf{a}}] = 
   \sum_{\sigma, \sigma'} [e^{xA}]_{\sigma, \sigma'} \delta_{\sigma, \sigma'}^{(\bc)} \,,
\begin{comment}
    \begin{cases}
        \sum_{\permstate} [e^{x A}]_{\permstate,\permstate} \;, & \mbox{pbc} \\
         \sum_{\permstate, \permstate'} [e^{x A}]_{\permstate,\permstate'} \;, & \mbox{obc}
    \end{cases}
\end{comment}
\end{equation}
where $\delta_{\sigma,\sigma'}^{(\bc)}$ reduces to Kronecker delta for pbc and to $1$ for obc and we used the normalization $\mathbb{E}[w] = 1$ to fix $\GinVar^2 = M^{-1}$~\footnote{For obc, we could get rid of the boundary states using that $\sum_{\alpha, \bar\alpha} \ell_{\alpha_1}\ell_{\bar\alpha_1}^\ast\ldots
\ell_{\alpha_k}l_{\bar\alpha_k}^\ast \llangle \alpha_1,\bar\alpha_1,\ldots, \alpha_k, \bar\alpha_k | \permstate \rrangle = |\ell|^{2k}$, which we can absorb in the definition of $\nu$.}.
The microscopic structure of the underlying circuit can only enter the scaling limit in setting the length scale $\Lth(t)$, while the general form of the moments only depends on the spectrum of the adjacency matrix $A$, for which we provide a proof in Appendixes~ \ref{app:SpecofA},\ref{app:gentoepltizFourier }. 
Expanding in powers of $x$, one sees that Eq.~\eqref{eq:momentsA} admits a simple interpretation as the grand canonical partition function of a dilute gas of domain walls, corresponding to transpositions connecting two permutations and each carrying a fugacity $\sim \Lth(t)^{-1}$ ({Fig.~\ref{fig:circuit}}(b)). Since a domain wall can be placed anywhere along the entire system, we obtain a factor $x= L/\Lth(t)$ for each of them. Finally, the composition of permutations and the boundary conditions impose selection rules on the allowed sequences of transpositions:  e.g., at the $n$-th order, for periodic conditions, only closed paths of length $n$ in the transposition graph are allowed, the number of which is given by $\tr{A^n}$.
In this perspective, the cost associated with an elementary transposition placed at position $\sim j$ can be identified with the membrane~\cite{jonay2018coarsegrained},  controlling the purity of $e^{-S_2(t)}$ of the subregion $[1,\ldots, j]$. Thus, we deduce $\Lth(t)^{-1} = \mathbb{E}[e^{-S_2(t)}]$ as anticipated, thus justifying the exponential growth of $\Lth(t)$ on general grounds.
It is worth commenting here on the case of Floquet random circuits in which the local gate $u_{j,j+1}(t')= u_{j,j+1}$ remains identical across different time steps, equivalent to quenched disorder. For sufficiently weak disorder, the dynamics still remain ergodic, but it is favourable to arrange domain walls at weak links. This mechanism is analogous to what was discussed in~\cite{PhysRevB.98.035118} and leads to predicting the same universal moments~\eqref{eq:momentsA} but a scaling $\log \Lth \propto t^\alpha$ with $\alpha<1$ and changing continuously with the strength of the disorder, in agreement with our numerical findings Appendix.~\ref{app:NumericalSim}.

As all nodes are equivalent in the transposition graph, the constant vector of $1$'s is the (maximal) eigenvector of $A$, with eigenvalue $k(k-1)/2$ given by the number of transpositions in $S_k$. Thus, Eq.~\eqref{eq:momentsA} implies
\begin{equation}
    \mathbb{E}[w^k_{\mathbf{a}}] \stackrel{\rm obc}{=} k! e^{x \nu(\rho \,=\, (k))} =k! e^{x k(k-1)/2} \,, 
    \label{eq:scalingobc}
\end{equation}
One easily recognises $e^{k(k-1)/2}$ as the moments of a lognormal distribution recovering Eq.~\eqref{eq:pobcw} via convolution. Interestingly, pbc are more involved as they require the knowledge of the full spectrum of $A$. As 
$A_{\sigma, \sigma'} = \delta_{d(\sigma, \sigma'), 1} = f(\sigma \sigma'^{-1})$, $A$
is an example of a matrix whose matrix elements only depend on the difference between the elements $\permstate, \permstate'$ in $S_k$. Also, $\forall \, \mu \in S_k, f(\mu \sigma \mu^{-1}) = f(\sigma)$, i.e. it depends only on the conjugacy class. We can see this as a generalization of a \textit{circulant Toeplitz matrix} and a generalized Fourier transform can diagonalise any such matrix. One has that its eigenvalues $\nu(\rho)$ are indexed by irreducible representations $\rho$ of $S_k$ with a degeneracy $\dim(\rho)^2$, $\dim(\rho)$ being the dimension of $\rho$ as explained in Appendix
~\ref{app:gentoepltizFourier }. Explicitly, we have 
\begin{equation}
    \label{eq:scalingpbc}
    \mathbb{E}[w^k_{\mathbf{a}}] \stackrel{\rm pbc}{=} \tr{e^{x A}} = \sum_{\rho\vdash k} \dim(\rho)^2 e^{x \nu(\rho)}  \,,
\end{equation}
where in the second line, we used that irreps of $S_k$ are labeled by integer partition of $k$. Since $\sum_{\rho\vdash k} \dim(\rho)^2=k!$, we recover the PT distribution for $x\to 0$. In~ Appendix~\ref{app:EQ2proof}, we show that these moments match those obtained from Eq.~\eqref{eq:pbssample}, which gives an effective procedure to compute the distribution of $w = w_0 g_x^{(pbc)}$.
\section{Random Phase Model (RPM)} To corroborate the universality derived in Eq.~\eqref{eq:momentsA}, let us consider the RPM~\cite{cdc2}. We consider
single-site Haar-random unitaries, $u_{i}^{(1)}(t')$, and  two-site gates, $[u_{j,j+1}^{(2)}(t')]_{a_j a_j+1, a_j a_j+1}=\exp[i\varphi^{(j)}_{a_j,a_{j+1}}(t')]$, coupling neighbouring sites via a diagonal random phase ($a_j = 0,2\ldots, q-1$). Each coefficient $\varphi_{a_j,a_{j+1}}^{(j)}(t')$ is an independent Gaussian random real variable with mean zero and variance $\epsilon$, which controls the coupling strength between neighbouring spins. Then, in the brick-wall geometry of Fig.~\ref{fig:circuit}(a), we choose
gates on even/odd layers as $u_{j,j+1}(t') = u^{(2)}_{j,j+1}(t') u_j^{(1)}(t')  u_{j+1}^{(1)}(t') $ or $u_{j,j+1}(t') = u_j^{(1)}(t')  u_{j+1}^{(1)}(t') u^{(2)}_{j,j+1}(t')$ respectively, so that all commuting $2-$site gates are applied one after the other. 
Constraining the gates $u^{(j)}(t')$ and $\varphi^{(j)}(t')$ to be site- or time-independent (or both), this model gives access to translational invariant and Floquet models, as explored in \cite{cdc2, cdclyap, chan2021manybody, Shivam_2023, huang2023outoftimeorder}. 
Here we first consider the case where all gates are drawn independently in space and time 
postponing the discussion of the Floquet case to the end. Also, we consider the analytically tractable limit $q\to\infty$ at fixed coupling $\epsilon$. To compute the moments of the overlap $\mathbb{E}[w^k_{\mathbf{a}}]$, we consider $k$ copies of the circuit and first consider the average over single-site random unitaries $u^{(1)}_j(t')$. Such an average can be once again expressed in terms of permutation states, using the formula $\mathbb{E}[u \otimes u^\ast \otimes \ldots u \otimes u^\ast ] = \sum_{\sigma, \tau \in S_k} \wg(\sigma \tau^{-1}) \kket{\sigma}\bbra{\tau} \stackrel{q\gg1}{\sim}q^{-k} \sum_{\sigma \in S_k} \kket{\sigma}\bbra{\sigma}$, where $\wg(\sigma)$ denotes the Weingarten function~\cite{Weingarten}. 
% permutation states arise upon ensemble averaging at each group of unitaries located at each space-time coordinate, and
Importantly, in contrast to Eq.~\eqref{eq:Gave},  permutation states arise upon ensemble averaging at each group of unitaries located at each space-time coordinate, and the contraction between permutation states occurs in the temporal direction for each spatial site, i.e. the vertical direction along a fixed site in Fig.~\ref{fig:circuit}(a). 
The decay of the overlap $\bbraket{\sigma | \sigma'} = q^{k - d(\sigma, \sigma')}$ when $q\to \infty$ forces the permutations to be the same in time for every fixed $j$. In other words, the Haar average at large $q$ leads to a sum over 
$k!^{L}$ possible permutations $\sigma_j$ at each spatial site $j$. The choice of permutations between neighbouring sites $j, j+1$ leads to different $\varphi^{(j)}$ random phase deletions. Specifically, the moments of the overlap can be expressed as
\begin{equation}
 \label{eq:wktrace}
    \mathbb{E}[w^k_{\mathbf{a}}]_{\rm RPM} = \; \sum_{\permstate_{1},\ldots, \permstate_{L} \in S_k} \prod_{j=1}^{L} [T_{\rm RPM}]_{\permstate_j,\permstate_{j+1}} \delta_{\permstate_1,\permstate_L}^{(\bc)} \,.
\end{equation}
where the transfer matrix $T_{\rm RPM}$ over a single site is obtained by averaging over all phases between two neighbouring sites. Using that the phases at different times are uncorrelated, we obtain $[T_{\rm RPM}]_{\sigma,\sigma'} =  \mathbb{E}[m_{\sigma, \sigma'}^{(k)}]^{t} $, where
the coefficient $m_{\sigma, \sigma'}^{(k)}$ is the contribution from the random phase gate in $u^{(2)}_{j,j+1}(t)$ at a given time slice
\begin{align}
m_{\sigma, \sigma'}^{(k)} = \; q^{-2 k}
   \sum_{\{a \}, \{b\} } 
   %\sum_{\substack{a_1,\ldots, a_k\\ b_1,\ldots, b_k}} 
   \prod_{j=1}^k e^{i \left[ \phi_{a_j, b_j} -  \, \phi_{a_{\sigma(j)}, b_{\sigma'(j)}} \right] }  \,,
\end{align}
where the summations runs over $a_i, b_i =0,1,\dots, q-1$, 
and contain repetitions when two or more indices $a_i, b_i$ have the same value. However, such coincidences can be ignored in the limit of large $q$, where we arrive at
\begin{equation}
    \mathbb{E}[m_{\sigma, \sigma'}^{(k)}]_{\mathrm{RPM}} = 
        e^{-\epsilon[k - \nf(\sigma \sigma^{\prime -1})]}  \;,
\end{equation}
where $\nf(\sigma)$ counts the number of fixed points in permutation $\sigma$.
With these expressions, we can now evaluate the moments in Eq.~\eqref{eq:wktrace} and the frame potential $F^{(k)}_{\rm RPM} = q^{-k L} \mathbb{E}[w^k]_{\rm RPM}$. 
In the limit of large $t$ at fixed $L$, the sum is dominated by the situation where all permutations are the same $\sigma_i = \sigma$ independently of $i$
and one recovers the PT distribution. Additionally, setting in this case $\Lth(t) = e^{2\epsilon t}$, we can expand
\begin{equation}
    [T_{\rm RPM}]_{\sigma, \sigma'} = \delta_{\sigma, \sigma'} + \frac{A_{\sigma, \sigma'}}{\Lth(t)} + O(\Lth(t))^{-2}
    \label{eq:TRPMexp}
\end{equation}
where $A$ is once again the adjacency matrix of the transposition graph. Thus, introducing the scaling variable $x = L/\Lth(t)$, from Eq.~\eqref{eq:wktrace} we recover Eq.~\eqref{eq:momentsA}. 
The RPM provides a useful framework for discussing the anticipated self-averaging property (as proven in Appendix~ \ref{app:self-aver}), demonstrating that the bitstring average of a single circuit realization coincides with the circuit average in the scaling limit~\cite{Note1}.

\section{Generalisation and discussion}
Unitary state design deals with sampling random Haar states with finite-depth unitary circuits. One expects convengence to Haar increasing the depth $t$ and the discrepancy can be estimated by looking at the overlaps between different realisations of circuits $W(t), W'(t)$: $\tilde w = \mathcal{N}|\braket{\Psi_0 | W^\dag(t) \tilde W'(t) | \Psi_0}|^2$. As the product  $W^\dag(t) W'(t)$ defines a circuit of depth $2t$, one has that $\tilde w$ follows the same universal distributions~\eqref{eq:pobcw} with $\Lth(2t)$.

The large $q$ analysis applied to RPM suggests that even for $d>1$ the moment $\mathbb{E}[w^k]$ is described by mapping to a model in $d$ dimensions in which local degrees of freedom are permutations and the ferromagnetic Boltzmann weight $J(t)$ between sites grows exponentially with $t$. At large times, there exist $k!$ perfectly ordered ground states. The excitations on top of these correspond to isolated defects~\cite{chan2021manybody}, in which a site differs by a single transposition from its neighbours. Since a defect can be placed anywhere in volume $N = L^d$ and its presence breaks $2d$ ferromagnetic bonds, the scaling limit in this case corresponds to $x = N/N_{\rm th}(t)$, with $N_{\rm th}(t) = J(t)^{2d}$. There are $\binom{k}{2}$ choices of transpositions for each defect and summing over their number, we arrive at $\mathbb{E}[w^k]_{d>1} = k! \sum_{n=0}^\infty (k(k-1)/2)^n x^n/n! = k! e^{k(k-1) x/2}$. This result coincides with Eq.~\eqref{eq:scalingobc} and leads again to Eq.~\eqref{eq:pobcw}, although its origin is different in that at $d>1$ the excitations are not domain walls but isolated defects. Numerical verification of this result is difficult, but recent quantum computing platforms offer a promising framework to observe our predictions, including higher $d>1$.

\section{Acknowledgements}
ADL thanks Adam Nahum and Tianci Zhou for discussions. AChr acknowledges support from the EUTOPIA PhD Co-tutelle program. AChan acknowledges support from the Royal Society grant RGS{$\backslash$}R1{$\backslash$}231444 and from the EPSRC Open Fellowship EP/X042812/1. We acknowledge support from the University of Warwick Library Open Access Fund. ADL acknowledges support from the ANR JCJC grant ANR-21-CE47-0003 (TamEnt). 
\nocite{CIT-006,diaconis1988,macdonald1998symmetric,doi:10.1142/S2010326319300018,10.1063/5.0048364,mac2019quantum}
\onecolumngrid
\newpage
\appendix
\section{Models}\label{app:models}
In this work, we focus on the quantum circuits with the brick-wall geometry as models of quantum many-body systems. Such models are defined with an evolution operator given by
\begin{align}\label{app_eq:bw_geometry}
	W(t)=\prod_{s=1}^t\tilde W(s),\quad 
	\tilde W(s)=\bigotimes_{\substack{j \in 2 \mathbb{Z} + s\, \text{mod}\,2}}u_{j,j+1}(s) \;.
\end{align}
For spatial-temporal random circuits, the two-site gates $u_{j,j+1}(s)$ are independent random variables drawn from the same ensemble for different $i$ and $s$. For Floquet model, $u_{j,j+1}$ are identical for different $s$. 

We consider 2 generic many-body quantum chaotic models below. 
For the \textit{brick-wall model} (BWM)~\cite{PhysRevX.8.021014}, $u_{i,i+1}(s)$ are independent random matrices drawn according to 
 \begin{align}
 	u^{\text{BWM}}_{j,j+1}(s)  \in \text{CUE}\left(q^2 \right) \, ,
\end{align}
where $\text{CUE}\left(n \right) $ is the circular unitary ensemble of unitaries of size $n$. For completeness, we repeat the definition of the random phase model here.  For the \textit{random phase model} (RPM)~\cite{cdc2},  
we consider single-site Haar-random unitaries, $u_{i}^{(1)}(t)$, and  two-site gates, $[u_{j,j+1}^{(2)}(t)]_{a_j a_j+1, a_j a_j+1}=\exp[i \varphi^{(j)}_{a_j,a_{j+1}}(t)]$, coupling neighbouring sites via a diagonal random phase ($a_j = 1,2\ldots, q$). Each coefficient $\varphi_{a_j,a_{j+1}}^{(j)}(t)$ is an independent Gaussian random real variable with mean zero and variance $\epsilon$, which controls the coupling strength between neighbouring spins. Then, in the brick-wall geometry, $u_{i,i+1}(s)$ are independent random matrices drawn according to
 \begin{align}
 	u^{\text{RPM}}_{j,j+1}(s) =
  \begin{cases}
  u^{(2)}_{j,j+1}(s)\, u_j^{(1)}(s)  \,u_{j+1}^{(1)}(s) \, ,   \qquad & 
  j \text{ even}
   \, ,
  \\
  u_j^{(1)}(t)  u_{j+1}^{(1)}(t) u^{(2)}_{j,j+1}(t) \,,
  \qquad & 
  j \text{ odd}
  \, ,
  \end{cases}
\end{align}
so that all commuting $2-$site gates are applied one after the other. 

\section{Spectrum of $A$}
\label{app:SpecofA}
As mentioned in the main text, the diagonalisation of $A$ is based on noting that the matrix elements $A_{\sigma, \sigma'} = f(\sigma \sigma^{\prime -1})$ for $\sigma, \sigma' \in S_k$ and $f: S_k \to \mathbb{R} $ a function dependent only on the conjugacy class. Any such matrix can be  diagonalised by a generalized Fourier transform~\cite{terras1999fourier}, $A = U \Lambda U^\dag$, where
\begin{equation} \label{eq:U_matrix}
    U_{\sigma,(\rho,ij)} = \sqrt{\frac{\mathrm{dim}(\rho)}{k!}} R_{\rho}(\sigma^{-1})_{ij} \,,
    %= \sqrt{\frac{\mathrm{dim}(\rho)}{k!}} \rho_{ji}(\sigma) \,,
\end{equation}
where $\rho$ labels the irreducible representations of $S_k$  and $R_\rho(\sigma)_{ij}$ are the components of the matrix representing the permutations $\sigma$ on $\rho$. The matrix $U$ is square and unitary as a consequence of the known identity $\sum_{\rho \in \Irr(S_k)} \dim(\rho)^2 = k!$.  The eigenvalues $\nu(\rho)$ are in one-to-one correspondence with  the irreps of $S_k$, with a degeneracy $\dim(\rho)^2$
corresponding to all choices of $i,j$. This clarifies the expansion in Eq.~\eqref{eq:scalingpbc}. 

Applying \eqref{eq:U_matrix}, one can express the eigenvalues $\nu(\rho)$ in terms of characters $\chi_{\rho}(\lambda) := \sum_{i} [R_{\rho}(\sigma)]_{ii}$ for any $\sigma \in \lambda$. 
We recall that, via Young diagrams, both the irreducible representations $\operatorname{Irr}(S_k)$ and the conjugacy classes $\operatorname{Cl}(S_k)$ are labelled by integer partitions of size $k$. 
We denote a conjugacy class $\mu \in \operatorname{Cl}(S_k) = (1^{a_1}2^{a_2} \dots k^{a_k})$ if any of its representatives is composed by $a_j$ $j$-cycles.
In general, one has the expression 
\begin{equation}
\nu(\rho) = 
\sum_{\mu \in \operatorname{Cl}(S_k)} f(\mu) \chi_{\rho}(\mu) \dim(\mu)/ \chi_{\rho}(1)
\end{equation}
which cannot be simplified further for a general $f$. 
In the case of $A$, the only conjugacy class that has non-zero $f(\mu)$ is by definition the $(1^{k-2} 2^1)$, and we recover
\begin{equation}
\label{eq:nudef}
\nu(\rho) = \binom{k}{2} \frac{\chi_\rho(1^{k-2}2^1)}{\chi_\rho(1^k)} = \frac 1 2\sum_i  \left[ \rho_i^2 - (\rho_i^\mathrm{t})^2  \right]  \;.
\end{equation}
and by $\rho^{\mathrm{t}}$ the dual partition, obtained by exchanging rows and columns in the corresponding Young diagram, i.e.  $\rho^\mathrm{t}_i = \# \{i | \rho_i \geq i\}$. For example, a partition of $k=4$ could be $(2,1,1)$ and its dual $(3,1)$ . Last equality is a consequence of the Frobenius formula~\cite{fulton2013representation}.

\section{Self-averaging property within RPM}\label{app:self-aver}
For a given realisation of the RPM, consider the $k$-th moment of the bitstring average $\mathcal{N}^{-1}\sum_{\textbf{a}} w_{\textbf{a}}^k =: \mathsf{m}^{(k)}_{W(t)}$ where we wrote as a subscript the dependence on the circuit. Then, its sample-to-sample variance
\begin{equation}
\label{eq:variance}
    \operatorname{Var}[\mathsf{m}^{(k)}_{W(t)}] = \mathbb{E}[(\mathsf{m}^{(k)}_{W(t)})^2] -
    \mathbb{E}[(\mathsf{m}^{(k)}_{W(t)})]^2  = 
   \left[\mathcal{N}^{-2} \sum_{\textbf{a}, \textbf{a}'} \mathbb{E}[w_{\textbf{a}}^k w_{\textbf{a}'}^k]\right] - \mathbb{E}[w^k_{\textbf{a}}]^2
\end{equation}

where the second term can be read squaring Eqs.~(\ref{eq:scalingobc}, \ref{eq:scalingpbc}). The first term, after circuit average and manipulations similar to those needed for Eq.~\eqref{eq:wktrace}, can be rewritten as (focusing on pbc for simplicity)
\begin{equation}
    \mathcal{N}^{-2} \sum_{\textbf{a}, \textbf{a}'} \mathbb{E}[w_{\textbf{a}}^k w_{\textbf{a}'}^k]  =  \Tr[ (T_{\rm RPM} V)^L] = \Tr\left[ \left(V + \frac{A V}{\Lth(t)} + O(\Lth(t)^{-2})\right)^L\right]  
\end{equation}
where the trace is over the space of permutations $\sigma \in S_{2k}$ and $V$ is a diagonal matrix with $V_{\sigma, \sigma} = 1$ if $\sigma$ is a factorised permutation $\in S_{k} \times S_k$ and $1/q$ otherwise. In the last equality we used Eq.~\eqref{eq:TRPMexp}. In large $\Lth(t)$, we can use perturbation theory in the degenerate subspace of factorized permutations. Then,  the $O(\Lth(t)^{-1})$ cancels completely in Eq.~\eqref{eq:variance}, so we are left with $\operatorname{Var}[\mathsf{m}^{(k)}_{W(t)}] = O(\Lth^{-1}(t))$, which vanishes in the scaling limit.

\section{Diagonalisation of the generalised circulant  Toeplitz matrix}\label{app:gentoepltizFourier }
%In this section, we first present the mathematical formalism that one can use to diagonalise a generalised Toeplitz matrix (to be defined later) and then apply it for the specific case of the adjacency  matrix $A$

%A $n \times n$ matrix $M$ is a Toeplitz matrix \cite{CIT-006} if  $M_{ij}=m(i-j)$ with $i,j=1,2,\dots,n$ for some function $m$. By definition, a Toeplitz matrix entry is defined by the relative position $i - j$, meaning the entries at each diagonal of the matrix are identical.
%\subsubsection{\blue{Toeplitz matrices of finite groups}}
 
 An $n$-by-$n$ matrix $M$ is a Toeplitz matrix \cite{CIT-006} if  $M_{ij}=m(i-j)$ with $i,j=1,2,\dots,n$ for some function $m$. As explained in the main text, given any function $f: \mathcal{G} \to \mathbb{C}$,
we can generalise the notion of the Toeplitz matrix to an arbitrary group $\mathcal{G}$, introducing a $|\mathcal{G}| \times |\mathcal{G}|$  matrix (with $|\mathcal{G}|$ the order of the group $\mathcal{G}$): 
\begin{equation}
\label{eq:gentoepl}
    F_{\sigma,\sigma'} = f(\sigma \sigma^{\prime -1})   \;, \qquad \forall \; \sigma,\sigma' \in \mathcal{G} \;. 
\end{equation}
Similarly to the case of standard Toeplitz matrices, the spectrum can be investigated using a generalised notion of the Fourier transform \cite{CIT-006}. Given a finite group $\mathcal{G}$, the group's representations $\rho:G \to \mathrm{GL}(d_\rho , \mathbb{C})$ with dimension $d_\rho$, and a function $f: \mathcal{G} \to \mathbb{C}$, we define its Fourier transform $\hat{f}(\rho)$  as a function over the space of representations of $\mathcal{G}$ which reads
\begin{equation}
\label{eq:ftdef}
    \hat{f}(\rho) = \sum_{\sigma \in \mathcal{G}} f(\sigma) \rho(\sigma) \;.
\end{equation}
The inverse of this relation can be shown  to be given by~ \cite{diaconis1988} 
\begin{equation}
    f(g) = \frac{1}{|\mathcal{G}|} \sum_{\rho\in \Irr(\mathcal{G})} \operatorname{dim}(\rho) \Tr[\rho(g^{-1}) \hat{f}(\rho)] \,, 
\end{equation}
where the sum is restricted to the irreducible representation $\Irr(\mathcal{G})$.
The nice property of this Fourier transform is that it converts convolutions into products. In other words for two functions $h, g: \mathcal{G} \to \mathbb{C}$, one gets
\begin{equation}
    h(\sigma) = \sum_{\sigma' \in \mathcal{G}}  f(\sigma \sigma^{\prime -1}) g(\sigma')  \quad \Rightarrow \quad \hat{h}(\rho) = \hat{f}(\rho) \hat{g}(\rho) \, .
\end{equation}
Now let's consider an eigenvector of the matrix $F$ in Eq.~\eqref{eq:gentoepl}. Labelling its components as $c(\sigma)$ for any $\sigma \in \mathcal{G}$, it must satisfy
\begin{equation}
    \sum_{\sigma' \in \mathcal{G}} f(\sigma \sigma^{\prime -1}) c(\sigma') = \lambda c(\sigma) \, .
\end{equation}
Taking the Fourier transform of both side, this implies
\begin{equation}
\label{eq:fteigen}
    \hat{f}(\rho) \hat{c}(\rho) = \lambda \hat{c}(\rho) \;, \qquad \forall \rho \in \Irr(\mathcal{G})
\end{equation}
Note that each side of this equation is a matrix of size $\dim(\rho)\times \dim(\rho)$. To solve this equation, let's write the spectral decomposition of the matrix $\hat{f}(\rho)$ in braket notation:
\begin{equation}
\label{eq:spectrf}
    \hat{f}(\rho) = \sum_{j=1}^{\dim(\rho)} \lambda_j(\rho) \ket{j}\bra{j}  \, .
\end{equation}
Then, we see that for any $\tilde{\rho} \in \Irr(\mathcal{G})$ and any pair $i,j  \in \{1,\ldots, \dim(\rho)\}$, the following choice of $\hat{c}(\rho)$ provides a solution of Eq.~\eqref{eq:fteigen}
\begin{equation}
    \hat{c}(\rho) \equiv \hat{c}^{(i,j,\tilde{\rho})} (\rho) = 
    \begin{cases}
      0  \, ,& \rho \neq \tilde{\rho} \,, \\
      \ket{i}\bra{j} \,, & \rho = \tilde{\rho} \,,
    \end{cases}
\end{equation}
where $\ket{i}$ and $\bra{j}$ refer respectively to the right and left eigenvectors of $\hat{f}(\rho)$.
Once plugged in Eq.~\eqref{eq:fteigen}, it leads to
\begin{equation}
    \hat{f}(\rho) \hat{c}^{(i,j,\tilde{\rho})}(\rho) = \lambda_i(\tilde{\rho}) \hat{c}^{(i,j,\tilde{\rho})}(\rho) \;.
\end{equation}
This shows that the spectrum of the matrix $F$ is given by the $\lambda_i(\rho)$ for $\rho \in \Irr(\mathcal{G})$ and $i = 1,\ldots,\dim(\rho)$ with a degeneracy $\dim(\rho)$, labeled by the index $j$. This provides a full spectral decomposition, since one has the known equality
\begin{equation}
    \sum_{\rho \in \Irr(\mathcal{G})} \dim(\rho)^2 = |\mathcal{G}|  \,.
\end{equation}
Now, let us consider the case where the function $f$ is a class function, 
i.e. it is invariant under the group conjugation
\begin{equation}
    f(\omega \sigma \omega^{-1}) = f(\sigma) \;,
\end{equation}
for every $\omega,\sigma \in \mathcal{G}$.
In this case, one can see that
\begin{equation}
    [\hat{f}(\rho), \rho(\sigma)] = 0  \;, \qquad \forall  \; \sigma \in \mathcal{G} \,.
\end{equation}
\newline
Indeed, by definition, we have
\begin{multline}
    \hat{f}(\rho) \rho(\sigma) = \sum_{\sigma' \in \mathcal{G}} f(\sigma') \rho(\sigma')\rho(\sigma) =
    \sum_{\sigma' \in \mathcal{G}} f(\sigma') \rho(\sigma' \sigma) 
    \sum_{\sigma'' \in \mathcal{G}} f(\sigma'' \sigma^{-1}) \rho(\sigma'') \\ = \sum_{\sigma'' \in \mathcal{G}} f(\sigma^{-1} \sigma'' ) \rho(\sigma'') = 
    \sum_{\sigma''' \in \mathcal{G}} f(\sigma''') \rho(\sigma \sigma''')  = \rho(\sigma) \sum_{\sigma''' \in \mathcal{G}} f(\sigma''') \rho( \sigma''')= 
    \rho(\sigma) \hat{f}(\rho) \,. 
\end{multline}
Because of Schur's lemma, if $\rho \in \Irr(\mathcal{G})$, $\hat{f}(\rho)$ must be a multiple of the identity
\begin{equation}
\label{eq:diagf}
    \hat{f}(\rho) = \lambda(\rho)\,  \mathbb{1} \,,
\end{equation}
and  therefore in the spectral decomposition Eq.~\eqref{eq:spectrf}, $\lambda_j(\rho) = \lambda(\rho)$ for all $j$'s.
For generalised Toeplitz matrices obtained by class functions, the eigenvalues are labelled by the irreducible representations $\rho$ and each has a degeneracy given by $\dim(\rho)^2$. We can finally obtain an equation for $\lambda(\rho)$ by taking the trace of both sides in Eq.~\eqref{eq:ftdef} and using \eqref{eq:diagf}
\begin{equation}
    \Tr[\hat{f}(\rho)] = \dim(\rho) \lambda(\rho) = \sum_{\sigma \in \mathcal{G}} f(\sigma) \chi_{\rho}(\sigma) \,, 
\end{equation}
where $\chi_{\rho}(\sigma) = \Tr[\rho(\sigma)]$ is the character of the representation $\rho$. Since both the function $f$ and the character are class functions, we can rewrite the sum as a sum over conjugacy classes $\operatorname{Cl}(\mathcal{G})$ 
\begin{equation}
\label{eq:eigenformula}
    \lambda(\rho) =  \sum_{\sigma \in \mathcal{G}} \frac{f(\sigma) \chi_{\rho}(\sigma)}{\chi_{\rho}(1)} = \sum_{\mu \in \operatorname{Cl}(\mathcal{G})} \frac{f(\mu) \chi_{\rho}(\mu) \dim(\mu)}{\chi_{\rho}(1)} \,, 
\end{equation}
where we used that $\chi_{\rho}(1) = \dim(\rho)$, since the representation of the neutral element is the  $dim(\rho)$ dimensional identity and we denote as $\dim(\mu)$ the size of the conjugacy class $\mu$.
As a consistency check, we can look at the trivial case where 
$f(\mu) = 1$ irrespectively of $\mu$. In this case, from Eq.~\eqref{eq:eigenformula}, we have
\begin{equation}
\label{eq:constf}
    \lambda(\rho) = \sum_{\sigma \in \mathcal{G}} \frac{\chi_{\rho}(\sigma)}{\chi_{\rho}(1)} = \delta_{\rho, 1} |\mathcal{G}| \,,
\end{equation}
where we indicate as $\rho = 1$ the trivial one-dimensional representation where all elements are sent to $1$. Eq.~\eqref{eq:constf} from the orthogonality of the characters
\begin{equation}
    \frac{1}{|\mathcal{G}|}\sum_{\sigma \in \mathcal{G}} \chi_{\rho}(\sigma) \chi_{\rho'}(\sigma) = \delta_{\rho,\rho'} \, ,
\end{equation}
choosing $\rho'=1$. Eq.~\eqref{eq:constf} is consistent with the fact that for $f = 1$, the matrix $F$ reduces to a matrix made of $1$'s, which thus has only one non-vanishing eigenvalue and which equals the size of the matrix itself, i.e. $|\mathcal{G}|$.

\section{Derivation of Eq.~\eqref{eq:pbssample}}
\label{app:EQ2proof}
We start using the standard results of \cite{BREZIN1996697, PhysRevE.58.7176} about the spectrum of a random matrix with an external deterministic source. Consider a matrix $M$ distributed according to %the measure
\begin{equation}
    \operatorname{Prob}(M) = \exp[-n \Tr[V(M) - A M]] \,, 
\end{equation}
where $V$ is the potential and $A$ is a deterministic matrix that we can assume to be diagonal without loss of generality $A= \mbox{diag}(a_1,\ldots,a_n)$. Then, the eigenvalues $\{w_1,\ldots, w_n\}$ of $M$ follow the joint probability distribution
\begin{equation}
\label{eq:brezinhikami}
\begin{aligned}
    &\operatorname{Prob}(w_1,\ldots, w_n)= \\ = \frac{1}{Z_n} \det(w_\alpha^{k-1})_{\alpha, k = 1}^n &\det(e^{n a_k w_\alpha})_{\alpha,k=1}^n \prod_{\alpha=1}^n e^{- n V(w_\alpha)} \,,
\end{aligned}
\end{equation}
where the constant $Z_n$ enforces normalization. For Eq.~\eqref{eq:pbssample}, one sets
\begin{equation}
M = \sqrt{x n} H + x D \,,
\end{equation}
where $H$ and $D$ are as defined in the main text, which is equivalent to choosing in Eq.~\eqref{eq:brezinhikami}
\begin{equation}
    V(M) = \frac{M^2}{2 x n} \;, \qquad A = \frac{D}{n} \,,
\end{equation}
leading to
\begin{equation}
\label{eq:prow}
\begin{aligned}
    &\operatorname{Prob}(w_1,\ldots, w_n) = \\ =\frac{1}{Z_n} \det(w_\alpha^{k-1})_{\alpha, k = 1}^n & \det(e^{- (k - 1/2) w_\alpha})_{\alpha,k=1}^n e^{-\sum_{\alpha=1}^n \frac{  w_\alpha^2}{2x}} \,.
\end{aligned}
\end{equation}
We are interested in computing the moments of $\Tr[e^{M}]$, i.e.
\begin{equation}
\begin{aligned}
  &  \Omega_k(x) := \left\langle \bigl(\sum_\alpha e^{w_\alpha}\bigr)^k \right\rangle \\= \int dw_1 & \ldots dw_n \operatorname{Prob}(w_1,\ldots, w_n) \left(\sum_\alpha e^{w_\alpha}\right)^k \,.
\end{aligned}
\end{equation}
The calculation will be analogous to~\cite{gerbino2024dyson}, but we report it here with the appropriate notation and normalizations for convenience. As a proxy for the calculation of $\Omega_k(x)$, we first introduce Schur's polynomials.
To an integer partition $\rho = (\rho_1,\ldots, \rho_n)$ of the integer $k = \sum_{j=1}^n \rho_j$, with
$\rho_1\geq\rho_2\geq\ldots\geq\rho_n \geq 0$, one associates
the corresponding Schur polynomial in $n$ variables $y_1,\ldots, y_n$ via~\cite{macdonald1998symmetric}
\begin{equation}
    s_\rho(y) := \frac{\det(y_\alpha^{\rho_j + n - j})_{j,\alpha=1}^n}{\det(y_\alpha^{k-1})_{k,\alpha=1}^n}= \frac{\det(y_\alpha^{h_j})_{j,\alpha=1}^n}{\det(y_\alpha^{k-1})_{k,\alpha=1}^n}
    \,, 
\end{equation}
where we denote $h_j \equiv \rho_j + n - j$. Schur polynomials are symmetric and homogeneous of degree $k$. Setting $y_\alpha = e^{w_\alpha}$ and using the Vandermonde determinant formula
\begin{equation}
\label{eq:vdm}
    \det(y_\alpha^{k-1})_{\alpha,k=1}^n = \prod_{\alpha<\beta} (y_\beta - y_\alpha) \;,
\end{equation}
We can deduce
\begin{equation}
    (-1)^{n(n-1)/2} e^{(n-1/2)\sum_\alpha w_\alpha} \det(e^{- (k - 1/2) w_\alpha})_{\alpha,k=1}^n = \det(e^{(k - 1) w_\alpha})_{\alpha,k=1}^n \, ,
\end{equation}
which allows us to express the average as
\begin{equation}
\label{eq:schurdet}
    \langle s_\rho(y = e^w) \rangle = \frac{(-1)^{n(n-1)/2}}{Z_n}  \int d w_1 \ldots d w_n \det(w_\alpha^{k-1})_{\alpha, k = 1}^n \det(e^{w_\alpha (h_j - n + 1/2)})_{j,\alpha=1}^n
    e^{-\sum_{\alpha=1}^n  \frac{  w_\alpha^2}{2x}} %e^{-(n-1/2)\sum_\alpha w_\alpha}
    \,.
\end{equation}
We can use Andreief identity~\cite{doi:10.1142/S2010326319300018} to express it in terms of a single determinant
\begin{equation}
\label{eq:andreief}
    \langle s_\rho(y) \rangle = 
    \frac{(-1)^{n(n-1)/2}
    (2\pi x)^{n/2} n!}{Z_n}
    \det( I_{k, \rho_j -j + 1/2})_{k,j=1}^n \;,
\end{equation}
where we defined
\begin{equation}
    I_{k,\ell} = \int_{-\infty}^\infty \frac{dw}{\sqrt{2 \pi x}} w^{k-1}  e^{ \ell w - \frac{w^2}{2x}} =  \left.\partial_{\mu}^{k-1} \left[ e^{x \mu^2/2} \right] \right|_{\mu = \ell} = \left(-i \sqrt{\frac{x}{2}} \right)^{k-1} e^{\ell^2 x/2} H_{k-1}\left(i \ell \sqrt{x/2} \right) \,,
\end{equation}
and in the last equality, we used the Hermite polynomials
$H_p(z) = (-1)^p e^{z^2} \partial_z^p [e^{-z^2}]$. Note that in these conventions, the leading coefficient is $H_p(z) = 2^p z^p + O(z^{p -1 })$. Thus, by using the properties of determinants, we can combine the rows to extract only the leading coefficient out of each Hermite polynomial, obtaining
\begin{equation}
\det[ I_{k, \rho_j -j + 1/2}]_{k,j=1}^n
= x^{n(n-1)/2} \exp\left[\frac x 2 \sum_j (\rho_j - j + 1/2)^2\right] \det[(\rho_j - j + 1/2)^{k-1}] \,.
\end{equation}
This last determinant is once again a Vandermonde one, which can be expressed via \eqref{eq:vdm}. We can now plug this back into Eq.~\eqref{eq:andreief} 
and fix the normalization $Z_n$ using that for the trivial partition of $0$, $\rho_1 = \rho_2 = \ldots \rho_n = 0$, so that $s_{\rho = 0}(y) = 1$ identically. We finally obtain
\begin{equation}
\label{eq:schurave}
\langle s_\rho(y) \rangle = 
\exp\left[\frac x 2 \sum_j (\rho_j - j + 1/2)^2 - (j + 1/2)^2\right]
s_\rho(1) \,,
\end{equation}
where we recognized the equality
\begin{equation}
\label{eq:schur1}
\prod_{1 \leq j<j' \leq n}\frac{\rho_j - \rho_{j'} - j  + j'}{j'-j}  = s_\rho(y_1 = 1, \ldots, y_n = 1) \,,
\end{equation}
which expresses the number of semi-standard Young diagram of shape $\rho$ and $n$ entries~\cite{macdonald1998symmetric}. Eq.~\eqref{eq:schurave} is consistent with the fact that for $x=0$, 
the distribution \eqref{eq:prow} reduces to $\operatorname{b}(w_1,\ldots, w_n) = \prod_\alpha \delta(w_\alpha)$
%$P(w_1,\ldots, w_n) = \prod_\alpha \delta(w_\alpha)$
as the matrix $M$ vanishes identically. Then, using the identity~\cite{macdonald1998symmetric}
\begin{equation}
    \sum_{j} (j-1)\rho_j = \frac 1 2\sum_i \rho_j^{\mathrm{t}} (\rho_j^{\mathrm{t}} -1) \,, 
\end{equation}
with $\rho^{\mathrm{t}}$ the dual partition of $\rho$, we obtain that
\begin{equation}
    \frac{1}{2}
     \sum_j (\rho_j - j + 1/2)^2 - (j + 1/2)^2 = \nu(\rho) \,,
\end{equation}
as defined in Eq.~\eqref{eq:nudef}.
Now, we can relate the average of Schur polynomials to $M_k(x)$ using (see Eq. 3.10 in \cite{10.1063/5.0048364})
\begin{equation}
    \Bigl(\sum_\alpha y_\alpha\Bigr)^k = \sum_{\rho \vdash k} \dim(\rho) s_\rho(y) \quad \Rightarrow \quad  M_k(x) = \sum_{\rho \vdash k} \dim(\rho) e^{x \nu(\rho)} s_{\rho}(1) \,.
\end{equation}
Finally, we consider the limit of large $n$. We have the standard identity~(see Example 5, page 46 in \cite{macdonald1998symmetric})
\begin{equation}
\lim_{n\to\infty}\frac{s_\rho(1)}{n^k}  = \frac{\dim(\rho)}{k!}\,, 
\end{equation}
which leads to the final result employed in the main text
\begin{equation}
    \mathbb{E}(g^k) = \lim_{n\to\infty} \frac{M_k(x)}{n^k} = \frac{1}{k!} \sum_{\rho \vdash k} \dim(\rho)^2 e^{x \nu(\rho)} \,.
\end{equation}

\section{Numerical simulations}
\label{app:NumericalSim}
\begin{figure}[t]
  \centering
    \begin{overpic}[width=0.45\linewidth]{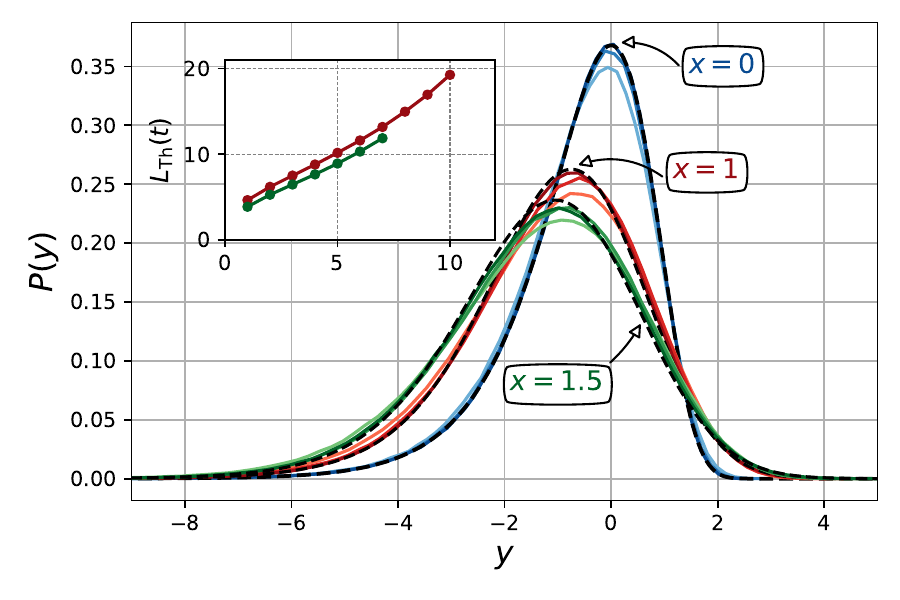}
      \put(53,-2){(a)}\put(18,20){\begin{tcolorbox}[
    colframe=black,          % Frame color (black)
    colback=gray!10,         % Background color (10% grey)
    rounded corners,         % Rounded corners for the box
    boxrule=0.5mm,           % Frame thickness
    width=0.1\linewidth,    % Width of the box
    enhanced,                % Enable advanced features
    opacityback=1,           % Full background opacity
    halign=center,           % Ensures text is horizontally centered
    valign=center,            % Ensures text is vertically centered
    overlay
]
RPM obc
\end{tcolorbox}}
    \end{overpic}
    \begin{overpic}[width=0.45\linewidth]{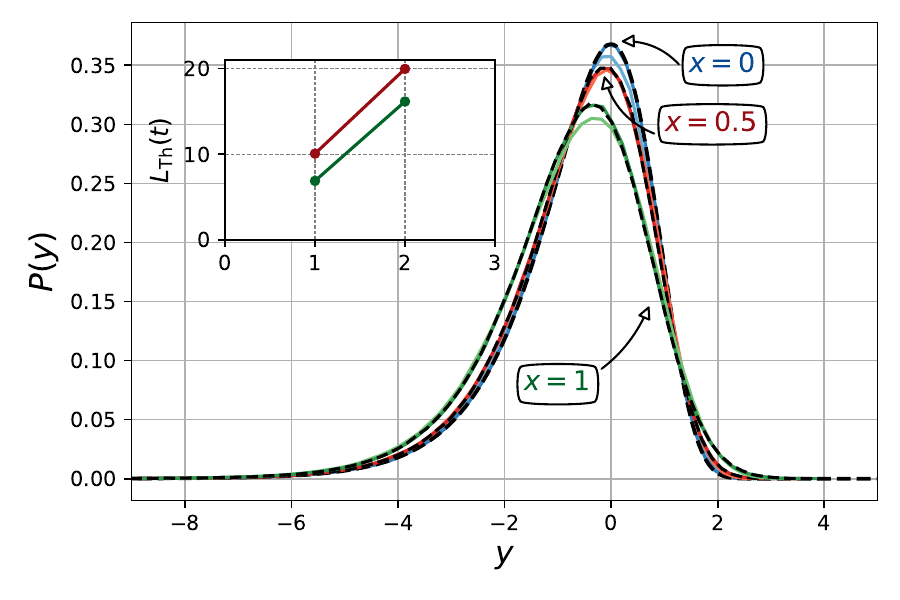}
      \put(53,-2){(b)} \put(18,18){\begin{tcolorbox}[
    colframe=black,          % Frame color (black)
    colback=gray!10,         % Background color (10% grey)
    rounded corners,         % Rounded corners for the box
    boxrule=0.5mm,           % Frame thickness
    width=0.1\linewidth,    % Width of the box
    enhanced,                % Enable advanced features
    opacityback=1,           % Full background opacity
    halign=center,           % Ensures text is horizontally centered
    valign=center,            % Ensures text is vertically centered
    overlay
]
BWM  pbc
\end{tcolorbox}}
    \end{overpic}
  \caption{Convergence of the numerical distributions (colored lines) to the theoretical ones (black-dashed line). (a): The obc numerical simulations of the RPM. For $x=0$, we provide data for $(t,L) \in \{ (10,6),(15,6),(20,6)\}$; for $x=1$, $(t,L) \in \{ (3,8),(5,10),(10,19)\}$; and for $x=1.5$, $(t,L) \in \{ (3,10),(5,13),(7,18)\}$. (b): Pbc numerical simulations for the BWM at $q=2$ and up to $L_{\text{max}}=20,\ t_{\text{max}}=20$. We provide data for $x=0$ at  $(t,L) \in \{ (3,15),(5,15)\}$; for $x=0.5$ at $(t,L) \in \{ (1,5),(2,10)\}$; for $x=1$ at $(t,L) \in \{ (1,7),(2,16)\}$.}
\label{fig:theoryconvercomplementary}
\end{figure}

\begin{figure}[H]
  \centering
    \begin{overpic}[width=0.45\linewidth]{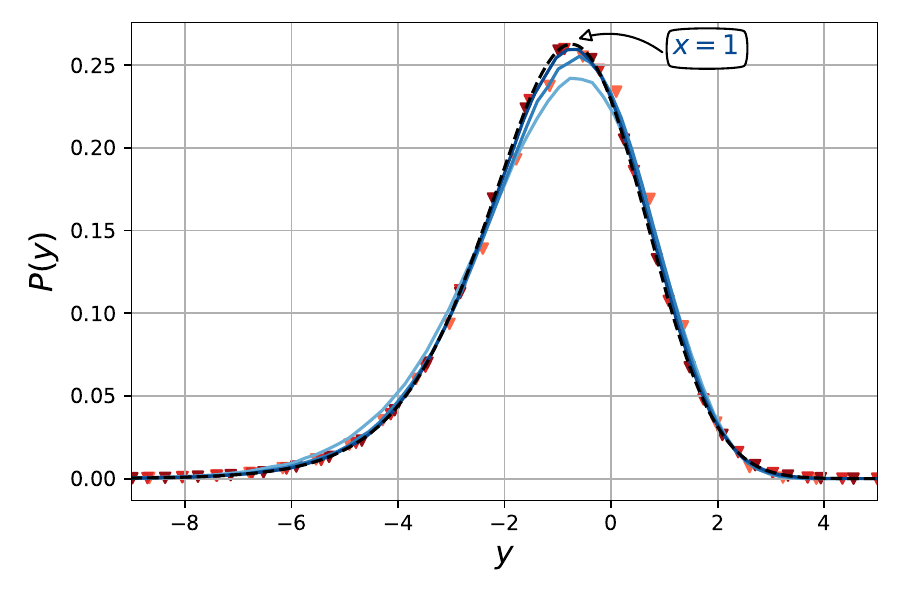}
      \put(53,-2){(a)}\put(18,40){\begin{tcolorbox}[
    colframe=black,          % Frame color (black)
    colback=gray!10,         % Background color (10% grey)
    rounded corners,         % Rounded corners for the box
    boxrule=0.5mm,           % Frame thickness
    width=0.15\linewidth,    % Width of the box
    enhanced,                % Enable advanced features
    opacityback=1,           % Full background opacity
    halign=center,           % Ensures text is horizontally centered
    valign=center,            % Ensures text is vertically centered
    overlay
]
RPM,BWM obc
\end{tcolorbox}}
    \end{overpic}
    \begin{overpic}[width=0.45\linewidth]{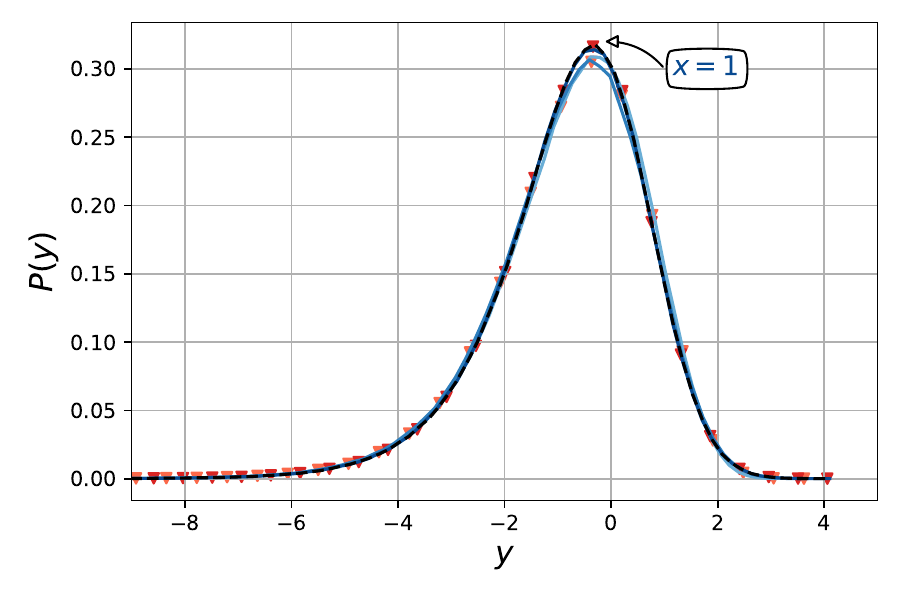}
      \put(53,-2){(b)}\put(18,40){\begin{tcolorbox}[
    colframe=black,          % Frame color (black)
    colback=gray!10,         % Background color (10% grey)
    rounded corners,         % Rounded corners for the box
    boxrule=0.5mm,           % Frame thickness
    width=0.15\linewidth,    % Width of the box
    enhanced,                % Enable advanced features
    opacityback=1,           % Full background opacity
    halign=center,           % Ensures text is horizontally centered
    valign=center,            % Ensures text is vertically centered
    overlay
]
RPM,BWM pbc
\end{tcolorbox}}
    \end{overpic}
  \caption{Convergence of both of the RPM (blue curves) and BWM (coloured triangles) models to the same scaling limit (black dashed curve) for  $x=1$. (a) The obc  numerical simulations of RPM at $(t,L) \in \{ (3,8),(5,10) ,(10,19)\}$ and of BWM at $(t,L) \in \{ (1,8),(3,40),(4,88)\}$; (b) The pbc numerical simulations of RPM at  $(t,L) \in \{ (3,6),(5,9),(10,17)\}$ and of BWM at $(t,L) \in \{ (1,7),(2,16)\}$.}
  \label{fig:bothconversclim}
\end{figure}

\begin{figure}[H]
    \centering
    \begin{overpic}[width=0.45\linewidth]
    {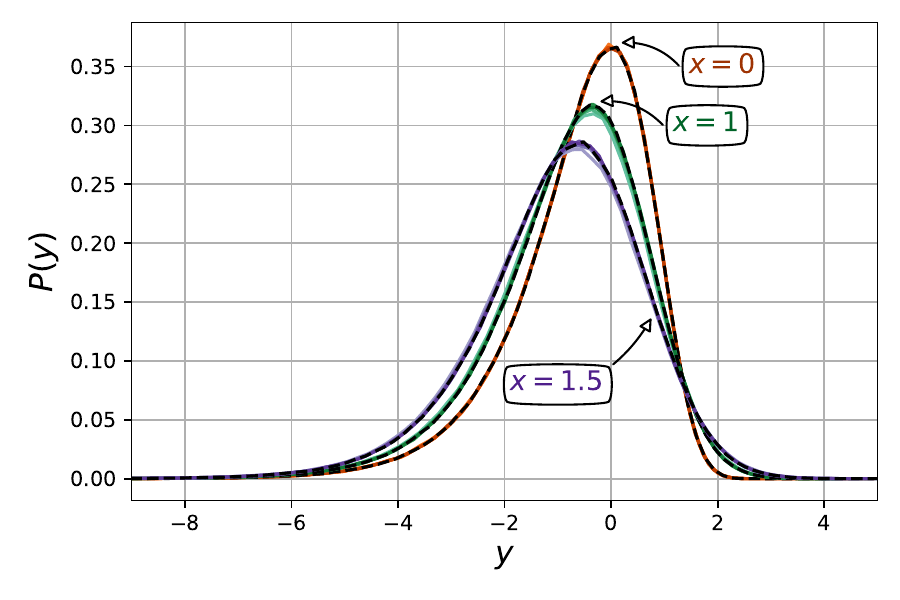}
      \put(53,-2){(a)} 
    \end{overpic}
    \begin{overpic}[width=0.45\linewidth]{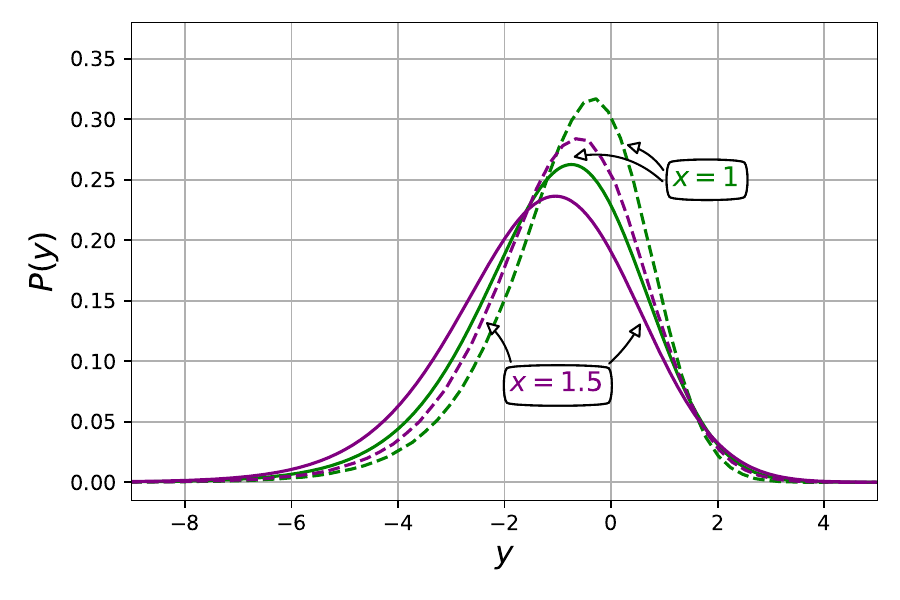}
      \put(53,-2){(b)}
    \end{overpic}
    \caption{(a) The convergence in $n$ for the  pbc theoretical prediction on $P(y)$ given by \eqref{eq:pbssample} (in the main text). The lines of the same colour correspond to $n=\{ 10,25,50,100,150\}$ from lighter to darker shade, with the black dashed line corresponding to $n=300$. (b)Comparison of the theoretical distributions for obc (solid curves) and pbc (dashed curves) at $x=1,1.5$.}
    \label{fig:nconvergence}
\end{figure}

%In this section we provide additional numerical results on the temporal-random and Floquet variations of two models, RPM  and BWM, defined in Appendix \ref{app:models}. 
%\subsection{Temporal-random circuits}
In the main text, we have demonstrated the convergence of the RPM, BWM in the pbc and obc cases. In this section, we further validate our findings by showcasing their consistency with the theoretical prediction in the complementary boundary conditions as demonstrated in Fig.~\ref{fig:theoryconvercomplementary}. Fig.~\ref{fig:bothconversclim} serves to explicitly confirm the universality of the Thouless scaling limit as predicted by our theoretical framework. The numerical  results in this paper were obtained in the following way:
\begin{itemize}
    \item \underline{\textit{RPM}}: The simulation was carried out in the time direction for systems up to maximum size $L_{\text{max}}=20$ and up to maximum time $t_{\text{max}}=20$ with an effective coupling strength $\epsilon=1$ and $q=2$. 
    % \brown{[Definitions requires use of dummy variables $s$ (instead of $t$). need to be consistent with the "Models" section. Don't need to explain the time direction simulation.]} In particular, we performed the dynamics by using the transfer matrix in the time direction. The single-time step transfer matrix is comprised of the two layers $W=W_{\text{odd}} W_{\text{even}}$ with $W_{\text{even/odd}}= \underset{j}{\otimes} \  u_{j,j+1}(t)$ (as detailed in the main text) and by applying it $t$ times,
    We computed the states of  $\ket{\Psi(t)}=W(t) \ket{\Psi_0}$, which in turn were used to generate the ensemble of $w=\mathcal{N}|\braket{\Psi|\Psi'}|^2$ for a sample size of 
    $N_{\text{sample}}=1.5 \times 10^6$. 
    %It is important to mention that the local two site gates $u_{j,j+1}(t)$ are chosen randomly at every site and time step. The Haar random unitaries, $u^{1}_j(t)$ and the invariance of the Haar measure under unitary operations, allow for a choice of the initial state, such that every local q-$q$it is at the same orbital  $\ket{\Psi_0}=\ket{0}^{\otimes L}$, without loss of generality.
    \item \underline{\textit{BWM}}: Similarly, we obtained the ensemble $w$ using the same methodology as for the RPM  with $q = 2$, $L_{\text{max}} = 20$, $t_{\text{max}} = 20$, but with employing the unitary circular ensemble (CUE) for the local gate $u_{j,j+1}(t)$. The BWM poses a greater numerical 
    challenge due to the rapid growth of $L_{\text{Th}}(t)$ over time, as observed in Fig.~\ref{fig:theoryconvercomplementary} and Fig.~\ref{fig:pdfbc} (in the main text). To address this, we employed the spatial transfer matrix method for simulating $\langle \Psi | \Psi' \rangle$ with $L_{\text{max}} = 120$ and $t_{\text{max}} = 5$. This method was specifically applied for the obc case, as the pbc scenario necessitates even more demanding computations, where we reached up to $t_{\text{max}} = 2$.
\end{itemize}
The theoretical distributions of the random variable $y=\log w$ were found using \eqref{eq:pobcw},\eqref{eq:pbssample} (in the main text) for obc and pbc, respectively. This analysis was carried out for $N_{\text{sample}}= 10^6$ at $x=0,0.5,1,1.5$. Fig.~\ref{fig:nconvergence}(a), illustrates that in the obc scenario, the distribution exhibits robust $ n-$ convergence at $n=300$ , which was utilised for numerical comparison. In Fig. ~\ref{fig:nconvergence}(b), we demonstrate the difference between the distributions for different boundary conditions.  The Thouless length
$L_{\text{Th}}(t)$ in our simulations was derived as $L_{\text{Th}}(t)=L_{\text{int}}(t)/x$. Here, $L_{\text{int}}(t)$ denotes the system size at which the numerical estimation of the average of $\mathbb{E}[y]_{\text{sim}}(L = L_{\text{int}}(t), t)$ matches the theoretical prediction $\mathbb{E}[y]_{\text{RPM/BWM}}$ for a specific time $t$ and value of $x$.
The fact that the $\Lth(t)$ length estimates obtained by this procedure give close values for different $x$ gives us strong confidence in the validity of the approach.

Here we also specify the parameters used in obtaining Fig. \ref{fig:pdfbc} in the main text. Specifically, in Fig. \ref{fig:pdfbc}(a), we present the pdf obtained from the pbc, a numerical simulation of the RPM at $q=2,\epsilon=1$. For $x=0$ and for which we demonstrate plots corresponding to the pairs $(t,L) \in \{(7,8), (11,8), (15,8)\}$; for $x=1$, $(t,L) \in \{(3,6), (5,9), (10,17)\}$; for $x=1.5$, 
$(t,L) \in \{(3,8), (5,11), (8,18)\}$. The theoretical distribution of $y$ was generated for $w=w_0 \, g$ using \eqref{eq:pbssample} and for a sample size $N_{\text{sample}}=10^6$ at $n=300$. In addition, in Fig. \ref{fig:pdfbc}(b), we showcase the pdf obtained from the obc numerical simulation for a brick wall model (BWM) where the local 2-site gate is chosen independently of the Haar distribution at $q=2$. The plots included correspond to data for $x=0$, $(t,L) \in \{(1,6), (3,6), (4,6)\}$; for $x=1$, $(t,L) \in \{(1,8), (3,40), (4, 88)\}$; for $x=1.5$, $(t, L) \in \{(1,11),(2,26), (3,57)\}$. The theoretical distribution $P(y)$ was created using \eqref{eq:pobcw}. All numerical distributions were obtained from a sample size $N_{\text{sample}}=1.5 \times  10^6$.


\begin{figure}[h]
  \centering
 \begin{minipage}[t]{0.45\textwidth}
 \includegraphics[width=\textwidth]{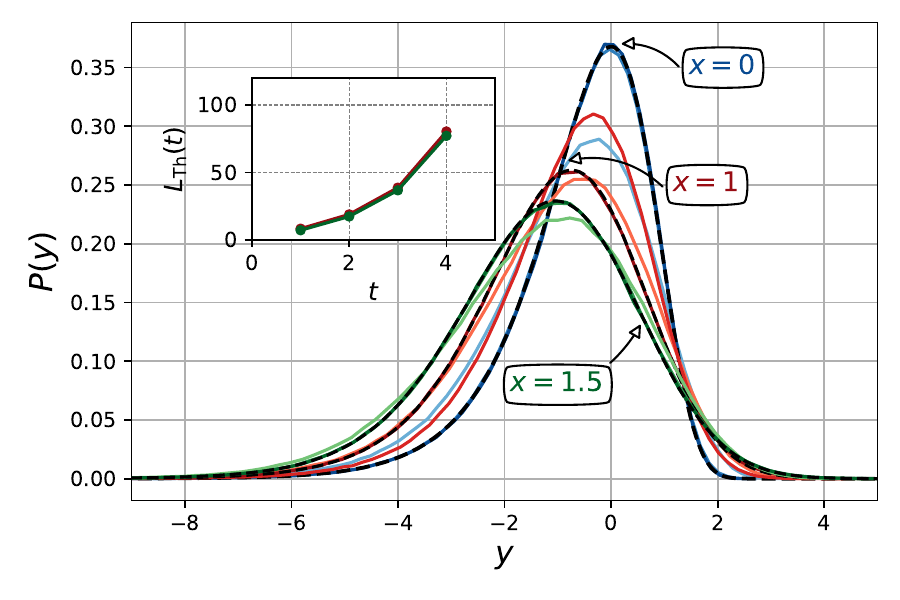}\put(-192,44){\begin{tcolorbox}[
    colframe=black,          % Frame color (black)
    colback=gray!10,         % Background color (10% grey)
    rounded corners,         % Rounded corners for the box
    boxrule=0.5mm,           % Frame thickness
    width=0.22\linewidth,    % Width of the box
    enhanced,                % Enable advanced features
    opacityback=1,           % Full background opacity
    halign=center,           % Ensures text is horizontally centered
    valign=center,            % Ensures text is vertically centered
    overlay
]
BWM obc
\end{tcolorbox}}
 \end{minipage}% 
  \hspace*{0.01\textwidth} 
 \begin{minipage}[t]{0.4519\textwidth}
 \centering
\includegraphics[width=\linewidth]{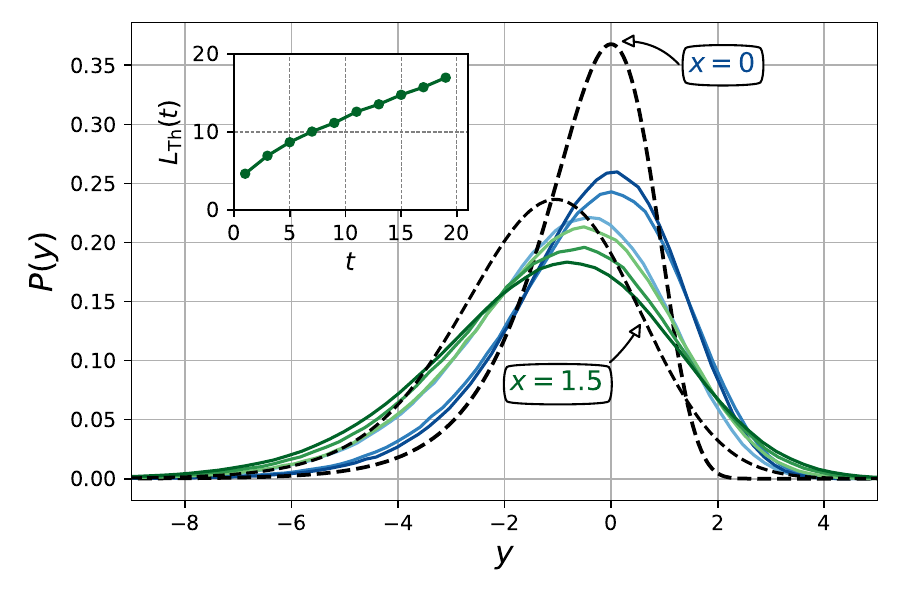}\put(-193,44){\begin{tcolorbox}[
    colframe=black,          % Frame color (black)
    colback=gray!10,         % Background color (10% grey)
    rounded corners,         % Rounded corners for the box
    boxrule=0.5mm,           % Frame thickness
    width=0.21\linewidth,    % Width of the box
    enhanced,                % Enable advanced features
    opacityback=1,           % Full background opacity
    halign=center,           % Ensures text is horizontally centered
    valign=center,            % Ensures text is vertically centered
    overlay
]
RPM obc
\end{tcolorbox}}
 \end{minipage}

  \caption{The numerical distributions for the Floquet circuit of a BWM (left) with obc, $q=2$  and of  RPM (right) with obc, $q=2,\epsilon=1$. In the former circuit, we observe agreement with the theory again and a fast-growing $L_{\text{Th}}(t)$, whereas in the latter one, that is not the case since the single-site random local unitaries induce a spatial disorder in the system, which is known to lead to a non-chaotic phase. The axes are the same as in Fig. \ref{fig:pdfbc}. }
    \label{fig:floquetcase}
\end{figure}

In the main text, we demonstrated the universality of the generalized PT distributions through simulations of two quantum many-body circuits. 
To provide further demonstration of the agreement, in Fig.~\ref{fig:floquetcase}, we simulate the Floquet BWM with obc at $q=2$, where the local gate $u_{j,j+1}(s)= u_{j,j+1}$ is independent of time step $s$. The specific parameters of the BWM are $x=0$ at  $(t,L) \in \{ (1,6),(3,6),(5,6)\}$; for $x=1$ at $(t,L) \in \{ (1,8),(3,19),(4,80)\}$; for $x=1.5$ at $(t,L) \in \{ (1,11),(3,55),(4,116)\}$.
To highlight the non-trivial nature of this agreement with theoretical predictions, in Fig.~\ref{fig:floquetcase}, we provide simulations of a Floquet quantum many-body MBL model, the RPM at $q=2$, which fails to exhibit the generalized PT distribution. This observation aligns with prior works \cite{cdc2, mac2019quantum}, which suggests that the $q=2$ Floquet RPM displays characteristics of a finite-size MBL phase, except at large effective coupling $\epsilon$. In this phase, we do not expect the Thouless length to grow unbounded and exponentially fast with time $t$ (see right inset of Fig.~\ref{fig:floquetcase}, thus invalidating the coarse-grained picture $\tilde{G}_a$ of the transfer matrix in the spatial direction. The specific parameters of the RPM are $q=2$, $\epsilon = 1$, for $x=0$ at  $(t,L) \in \{ (5,8),(10,8),(20,8)\}$; for $x=1.5$ at $(t,L) \in \{ (11,11),(13,14),(15,17)\}$.

\begin{comment}
\subsection{Floquet circuits}
Here we present our numerical findings concerning Floquet circuits, where the local gate $u_{j,j+1}(s)= u_{j,j+1}$ are independent of time step $s$ for both the RPM and BWM models. Fig.~\ref{fig:floquetcase} illustrates the convergence of the Floquet BWM to our theoretical predictions, while the Floquet RPM at $q=2$ does not show convergence to our theoretical predictions. This observation aligns with prior works \cite{cdc2, mac2019quantum}, which suggests that the $q=2$ Floquet RPM displays characteristics of a many-body localized (MBL) phase, except at large effective coupling $\epsilon$. In the MBL, we do not expect the Thouless length to grow unbounded and exponentially fast with time $t$ (see inset of Fig.~\ref{fig:floquetcase}(a)), thus invalidating the coarse-grained picture $\tilde{G}_a$ of the transfer matrix in the spatial direction.
\end{comment}

\twocolumngrid \bibliography{biblio}
\end{document}